\documentclass[
    aps,prd,
    twocolumn,
    nofootinbib,
    superscriptaddress,
]{revtex4}
\usepackage[utf8]{inputenc}
\usepackage{lmodern}
\usepackage{amsmath}
\usepackage{amssymb}
\usepackage{makecell}
\usepackage{amsfonts,bm}
\usepackage{xspace}
\usepackage{graphicx}
\usepackage{float} 
\usepackage{paralist}
\usepackage[colorlinks]{hyperref}
\usepackage{xcolor}
\usepackage[normalem]{ulem}

\newcommand{\cg}{\ensuremath{c_{\gamma}^{\text{eff}}}\xspace}
\newcommand{\thet}{\ensuremath{\theta_{\eta}}\xspace}

\begin{document}

\title{Phenomenology of axion-like particles with universal fermion couplings -- revisited}

\hfill \parbox{17.6cm}{\flushright TTP23-042, P3H-23-070 \vspace{1.5ex}}

\author{Giovani~Dalla~Valle~Garcia}
\email{giovani.garcia@student.kit.edu}
\affiliation{Institut für Astroteilchen Physik, Karlsruher Institut für Technologie (KIT), Hermann-von-Helmholtz-Platz 1, 76344 Eggenstein-Leopoldshafen, Germany} 
\author{Felix~Kahlhoefer}
\email{kahlhoefer@kit.edu}
\affiliation{Institute for Theoretical Particle Physics (TTP), Karlsruhe Institute of Technology (KIT), D-76131 Karlsruhe, Germany}
\author{Maksym~Ovchynnikov}
\email{maksym.ovchynnikov@kit.edu}
\affiliation{Institut für Astroteilchen Physik, Karlsruher Institut für Technologie (KIT), Hermann-von-Helmholtz-Platz 1, 76344 Eggenstein-Leopoldshafen, Germany}
\affiliation{Instituut-Lorentz, Leiden University, Niels Bohrweg 2, 2333 CA Leiden, The Netherlands}
\author{Andrii~Zaporozhchenko}
\email{andriizaporozhchenko@knu.ua}
\affiliation{Taras Shevchenko National University of Kyiv, 64 Volodymyrs’ka str., Kyiv 01601, Ukraine}

\date{\today}

\begin{abstract}

Axion-like particles (ALPs) emerge in many extensions of the Standard Model as pseudo-Goldstone bosons of a spontaneously broken global symmetry. Understanding their phenomenology in high-energy collisions is crucial for optimizing experimental searches and understanding the exploration potential of future experiments. In this paper, we revise the phenomenology of ALPs with universal couplings to fermions. In particular, we analyze the hierarchy and uncertainty of the various ALP production channels depending on the proton collision energy and the placement of the experiment, and provide improved calculations of the hadronic decay modes.

\end{abstract}

\maketitle

\section{Introduction}
\label{sec:introduction}

Axion-like particles (ALPs) $a$ are pseudoscalar particles that arise in theories with spontaneously broken global chiral symmetries, generalizing the idea of the QCD axion -- a hypothetical light particle capable of solving the strong CP problem~\cite{Peccei:1977hh,Weinberg:1977ma,Wilczek:1977pj}. While the QCD axion obtains its mass directly from its coupling to gluons, a generic ALP may interact with various particles and have an arbitrary mass~\cite{Jaeckel:2010ni}. The lowest-order gauge-invariant Lagrangian of the ALP interaction takes the form~\cite{Georgi:1986df,Bauer:2017ris,Brivio:2017ije}
\begin{align}
    \mathcal{L} =  \frac{a}{F}\bigg( & C_{G}\frac{\alpha_{s}}{4\pi}G_{\mu\nu}^{c}
    \tilde{G}^{\mu\nu,c} + C_{W}\frac{\alpha_{W}}{4\pi}W^{\mu\nu,c}\tilde{W}^{c}_{\mu\nu} \nonumber \\ & + C_{B}\frac{\alpha_{B}}{4\pi}B_{\mu\nu}\tilde{B}^{\mu\nu}\bigg) + \frac{\partial^\mu a}{F} \sum_F \bar{\Psi}_F \, \mathcal{C}_F\, \gamma_\mu\, \Psi_F  \; ,
    \label{eq:model-gauge}
\end{align}
where $G,W,B$ are field strengths of the Standard Model $SU_{c}(3)$, $SU_{L}(2)$, and $U_{Y}(1)$ gauge groups, $\alpha_{s},\alpha_{W},\alpha_{B} = g^{2}/4\pi$ are corresponding running couplings, $\tilde{G}_{\mu\nu} = \frac{1}{2}\epsilon_{\mu\nu\alpha\beta}G^{\alpha\beta}$ is the dual strength and $\Psi_F$ denote the SM fermion multiplets. Furthermore, $F$ is a dimensional scale, $C_{G,W,B}$ are dimensionless parameters, and $\mathcal{C}_F$ are hermitian matrices characterizing the structure of the ALP couplings to fermions. For light ALPs with $m_{a} \simeq 1\text{ GeV}$, past experiments have excluded  combinations $F/C_{i} \lesssim 1\text{ TeV}$~\cite{Beacham:2019nyx,Antel:2023hkf}. Therefore, ALPs belong to the class of the so-called Feebly-Interacting Particles, or FIPs. Many studies have explored the phenomenology of the individual terms in the Lagrangian above, in particular couplings to gluons~\cite{Aloni:2018vki,Aloni:2019ruo,Chakraborty:2021wda}, photons~\cite{Dobrich:2015jyk,Dolan:2017osp,Dobrich:2019dxc}, electroweak gauge bosons~\cite{Izaguirre:2016dfi,Alonso-Alvarez:2018irt,Gavela:2019wzg}, fermions~\cite{Dolan:2014ska,Dobrich:2018jyi,Carmona:2021seb}, the effect of flavour-violation~\cite{Cornella:2019uxs,Calibbi:2020jvd} and renormalisation group evolution~\cite{Chala:2020wvs,Bauer:2020jbp,Bauer:2021mvw,Ferber:2022rsf}, as well as the interplay between different couplings~\cite{Ertas:2020xcc,Jerhot:2022chi,Liu:2023bby,Bruggisser:2023npd}.

One case of particular interest is ALPs that interact dominantly with fermions with universal and flavour-diagonal couplings,
\begin{equation}
    \mathcal{L}_{\text{eff}} = \frac{\partial_{\mu}a}{F}\left(C_{\ell}\sum_{\ell}\bar{\ell}\gamma^{\mu}\gamma_{5}\ell +C_{q}\sum_{q}\bar{q}\gamma^{\mu}\gamma_{5}q \right)
    \label{eq:model-fermions-1}
\end{equation}
where $\ell$ are leptons and $q$ are quarks~\cite{Quevillon:2019zrd,Arias-Aragon:2022iwl}. The case $C_{\ell} = C_{q}$ has been identified by the Physics Beyond Colliders (PBC) initiative~\cite{Beacham:2019nyx} as one of the benchmark models (called BC10) to demonstrate the FIP exploration potential of future experiments. Due to a lack of in-depth theoretical studies of this model, the description of the phenomenology of such ALPs proposed in Ref.~\cite{Beacham:2019nyx} suffers from two issues. First, following Ref.~\cite{Dolan:2014ska}, the contribution of the hadronic decays to the total ALP decay width is assumed to be negligible. While this is a reasonable assumption for light ALPs with $m_{a} \ll 1\text{ GeV}$, where the only relevant decay mode is into three pions~\cite{Bauer:2020jbp}, the hadronic decays may actually dominate the total width for heavier ALPs, as indicated by calculations of the ALP decay width into quark and gluon pairs~\cite{Bauer:2020jbp,Ferber:2022rsf}. 

Another problem is that the production of such ALPs at beam dump experiments has been approximated by considering decays $B\to X_{s}+a$, where $X_{s} = K,K^{*}$ only~\cite{Dobrich:2018jyi,Ferber:2022rsf}. Nevertheless, there may be a sizable contribution from the $B$ decays into other resonances, $X_{s} = K_{1}, K^{*}_{2}, K^{*}_{0}$. For the case of light Higgs-like scalars~\cite{Boiarska:2019jym}, which couple to the $b\to s$ operator in a way similar to ALPs, these decays have been shown to correspond to $1/3$ of all possible decays. The same effect can be expected to be relevant for the case of ALPs with interactions as in eqs.~\eqref{eq:model-gauge} and~\eqref{eq:model-fermions}. Second, additional production processes arise due to the mixing of the ALPs with light pseudoscalar mesons $m^{0} = \pi^{0}/\eta/\eta'$, similar to the ALPs coupled to gluons.

In this paper, we address these issues by revising the phenomenology of the model given in eq.~\eqref{eq:model-fermions}. Our goal is to provide a detailed and comprehensive description of the PBC BC10, which the community may easily implement to consistently interpret ALP constraints from existing experiments and study the projected sensitivities of proposed searches. The results summarized in this paper are also accessible in a \texttt{Mathematica} notebook supplementing the paper (see Appendix~\ref{app:mathematica-notebook}).\footnote{Available on \href{https://github.com/maksymovchynnikov/ALPs-phenomenology}{https://github.com/maksymovchynnikov/ALPs-phenomenology}} We implement the revised phenomenology in \texttt{SensCalc}~\cite{Ovchynnikov:2023cry} -- a public code that uses the semi-analytic approach to calculate the number of events with decays of FIPs and sensitivities of proposed experiments.

The remainder of this work is organized as follows. In Sec.~\ref{sec:model-details}, we discuss the details of the ALP model that we consider, and in particular, the choice of the scale at which the underlying Lagrangian is defined. In Sec.~\ref{sec:production}, we discuss various contributions to the ALP production flux depending on the proton collision energy for the given experiment and its geometric placement. In Sec.~\ref{sec:decays}, we study the decay palette of the ALP. We conclude in Sec.~\ref{sec:conclusions}.

\section{Model details}
\label{sec:model-details}

The phenomenology of GeV-scale ALPs depends on the scale $\Lambda$ at which the interactions in eqs.~\eqref{eq:model-gauge},~\eqref{eq:model-fermions-1} are defined. This is because of a non-trivial renormalization group (RG) flow, which generates additional effective couplings absent in eqs.~\eqref{eq:model-gauge},~\eqref{eq:model-fermions-1} and modifies the initial couplings $C_{i}$ if the energy scale $\mu$ of the process under consideration differs from $\Lambda$. 

Typically, $\Lambda$ and $F$ are closely related, $F\simeq \Lambda$~\cite{Bauer:2020jbp}. The values of the Wilson coefficients $C_{i}(\Lambda)$ at that scale, however, are in general unknown, so the quantities $f_{i} \equiv F/C_{i}(\Lambda)$, controlling the interaction strength of ALPs with the corresponding SM fields, can be treated as independent parameters. Here and below, we will introduce the coefficients $c_{i}(\mu) \equiv C_{i}(\mu)/C_{i}(\Lambda)$, such that by construction $c_{i}(\Lambda) \equiv 1$, and perform our studies in terms of the effective suppression scales $f_i$, assuming that $f_i > 0$ for definiteness. The assumption of universal couplings to all the fermions then corresponds to a single scale $f_f = f$. With this assumption and setting $\Lambda = 1\text{ TeV}$, the model~\eqref{eq:model-fermions-1} exactly coincides with the BC10 model from~\cite{Beacham:2019nyx}, such that our results may be directly used by the experimental community:
\begin{equation}
    \mathcal{L}_{\text{eff}} = \frac{\partial_{\mu}a}{f}\left(c_{\ell}\sum_{\ell}\bar{\ell}\gamma^{\mu}\gamma_{5}\ell +c_{q}\sum_{q}\bar{q}\gamma^{\mu}\gamma_{5}q \right).
    \label{eq:model-fermions}
\end{equation}
For a specific ultraviolet model that predicts the Wilson coefficients $C_f(\Lambda)$, our results can be directly re-interpreted in terms of the fundamental scale $F$.

The RG evolution has been thoroughly studied for general models of ALPs (see Refs.~\cite{Bauer:2021mvw,Bauer:2020jbp,Ferber:2022rsf} and references therein). The evolution is usually split onto the flow from $\Lambda$ down to the electroweak scale $\mu_{w} \equiv m_{t}$ and from $\mu_{w}$ down to the scale of the process with ALPs, which is of the order of the ALP mass. For the processes with hadrons, there is one additional scale $\sim 4\pi\Lambda_{\text{QCD}}$ where the perturbative QCD should be matched with Chiral Perturbation Theory (ChPT).

\begin{figure*}
    \centering
    \includegraphics[width=0.43\textwidth]{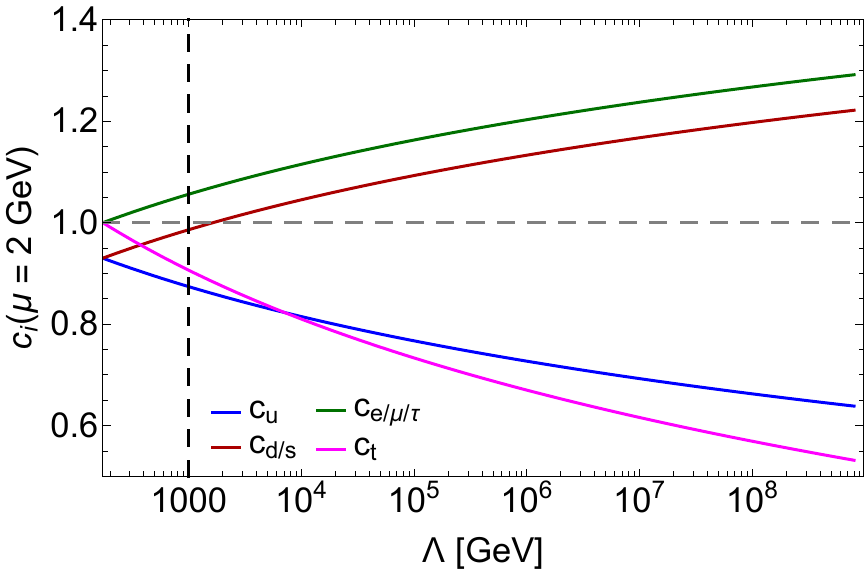}~ \includegraphics[width=0.45\textwidth]{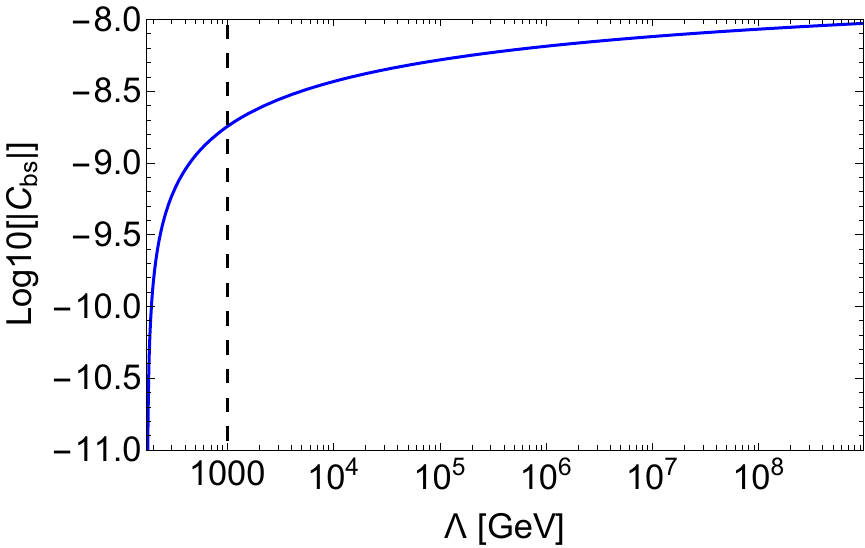}
    \caption{The RG flow of the quark and lepton couplings $c_{\ell},c_{q}$ from the Lagrangian~\eqref{eq:model-fermions} (the left panel) and modulus of the flavor-changing $abs$ coupling from eq.~\eqref{eq:effective-fcnc} at $f = 1\text{ PeV}$ (the right panel). The scale $\Lambda$, at which the interactions~\eqref{eq:model-fermions} are defined (and where, by construction, $c_{f}(\Lambda) \equiv 1$), is assumed to be between $m_{t} \approx 173\text{ GeV}$ and $10^{9}\text{ GeV}$. The values of the couplings are shown for the scale $\mu = 2 \text{ GeV}$. The vertical dashed line shows the reference scale $\Lambda = 1\text{ TeV}$ used as our benchmark choice.}
    \label{fig:RG-flow}
\end{figure*}

The couplings $C_{G}, C_{W}, C_{B}$ entering the Lagrangian in eq.~\eqref{eq:model-gauge} are invariant under the RG evolution at least up to the second order in the loop expansion. Therefore, they will only be generated from the initial Lagrangian in eq.~\eqref{eq:model-fermions} through threshold corrections when integrating out heavy quarks. This is not the case for the fermion couplings from eq.~\eqref{eq:model-fermions}, which may also evolve due to electroweak and strong interaction loops. To study the dynamics of these couplings, we solve the RG equations from Ref.~\cite{Bauer:2020jbp}. As for the couplings to heavy quarks $q = c,b,t$, we define that they do not run below the scale $\mu = m_{q}$ as the corresponding degrees of freedom are integrated out; in particular, $c_{t}(\mu<\mu_{w}) = c_{t}(\mu_{w})$. In Fig.~\ref{fig:RG-flow} (left panel), we show the value of the running couplings $c_{i}(\mu)$ from eq.~\eqref{eq:model-fermions} to various SM fermions at the scale $\mu = 2\text{ GeV}$ as a function of $\Lambda$.

The RG dynamics of the lepton couplings $c_{\ell}$ is driven by electroweak interactions. The ratio $(c_{\ell}(\Lambda)-c_{\ell}(\mu))/c_{\ell}(\Lambda)$ is $\lesssim 0.2$ for $\Lambda \lesssim \mathcal{O}(10\text{ TeV})$. Since all the leptons have the same properties under the electroweak gauge group (such as the weak isospin and electric charge), the flow of $c_{\ell}$ is lepton-flavor independent. 

This is not true for the quark couplings. Their evolution down to $\mu_{w}$ is flavor universality violating because different quarks have different weak isospins and electric charges. Assuming $\Lambda = 1\text{ TeV}$, the relative difference between $u$ and $d,s$ couplings is $(c_{d/s}(\mu)-c_{u}(\mu))/c_{d/s}(\mu) \simeq 10\%$. As we will see in Sec.~\ref{sec:production}, this difference cannot be neglected when studying the interactions of the ALPs with neutral pions (see Sec.~\ref{sec:production}). 

The coupling universality between quarks and leptons gets broken even if assuming $\Lambda = \mu_{w}$. This is because below $\mu_{w}$ the flow of the $c_{q}$s is determined by loops involving strong interactions, resulting in the corrections of the order of $10^{-2}c_{q}$, whereas $c_{\ell}$s evolve only due to the EM corrections, which are of the order of $10^{-5}c_{\ell}$~\cite{Bauer:2021mvw}.

The ALP-gluon interaction is another type of interaction important for ALP production and decay. At the leading order in $\alpha_{s}$, the matrix element of the type $GG\to a$ (the gluon fusion process) or $a\to GG$ (the ALP decay into a pair of gluons) is generated by the following matrix element:
\begin{equation}
   \mathcal{M}_{GG\leftrightarrow a} \approx c^{\text{eff}}_{G}(m_{a})\frac{\alpha_{s}}{2\pi f}G^{a}_{\mu\nu}(p_{G,1})\tilde{G}^{\mu\nu,a}(p_{G,2}),
    \label{eq:effective-gluon}
\end{equation}
where $G^{a}_{\mu\nu}(p)$ is the linear part of the gluon strength tensor with the replacements $\partial_{\mu}\to ip_{\mu}$  and $G_{\mu}^{a}\to \epsilon_{\mu}^{a}(p)$, with $\epsilon_{\mu}^{a}$ being the polarization vector. The effective coupling $c^{\text{eff}}_{G}(m_{a})$ is~\cite{Bauer:2020jbp}
\begin{equation}
     c^{\text{eff}}_{G} \equiv \sum_{q}c_{q}B_{1}\left( \frac{4m_{q}^{2}}{m_{a}^{2}}\right)
    \label{eq:cGeff}
\end{equation}
with 
\begin{equation}
    B_{1} = 1 - \tau \times \begin{cases} \arcsin^{2}\left(\frac{1}{\sqrt{\tau}}\right), \quad \tau \geq 1  \\ \left( \frac{\pi}{2}+\frac{i}{2}\ln\left[\frac{1+\sqrt{1-\tau}}{1-\sqrt{1-\tau}} \right]\right)^{2}, \quad \tau < 1\end{cases} \, .
    \label{eq:B1}
\end{equation} 
and $\tau_{f} = 4m_{f}^{2}/m_{a}^{2}$. In the regime $\tau \gg 1$, the function $B_{1}$ behaves as $B_{1}(\tau \gg 1) \approx -(3\tau)^{-1}$. In the opposite regime, $B_{1}(\tau \ll 1)\approx 1$. Therefore, the coupling to gluons is mainly generated by the quarks lighter than the ALP.

Additionally, the RG flow (via loops involving top quarks and charged weak gauge bosons) generates the flavor-changing neutral current (FCNC) coupling $D_{i}D_{j}a$, where $D = d,s,b$ are down quarks and $i\neq j$~\cite{Bauer:2021mvw}:
\begin{equation}
    \mathcal{L}_{\text{FCNC}} =  \frac{ \partial_{\mu} a}{f}\sum_{i\neq j} c_{ij} \, \bar{q}_{i}  \gamma^{\mu} P_{L} q_{j}
    \label{eq:effective-fcnc-0}
\end{equation} 
where $P_{L} = \frac{1}{2}(1-\gamma_{5})$ and $c_{ij}$ is the model-dependent coupling:
\begin{align} 
    c_{ij} = - V_{ti}^{\ast} V_{tj} \Big\{ & -\frac{1}{6}I_{t}(\Lambda, \mu_{w}) \nonumber \\ + \dfrac{y_t^2}{16\pi^2} & \bigg[c_{q} \bigg(
\ln{\dfrac{\mu_w^2}{m_t^2}}+ \frac{1}{2} + 3 \dfrac{1 - x_t + \ln{x_t}}{(1-x_t)^2}   \bigg) \nonumber \\ & + \dfrac{9\alpha_{\text{EM}}}{4 \pi s_w^2}  \left(  3 c_q + \ell \right)  \dfrac{1 - x_t + x_t\ln{x_t}}{(1-x_t)^2}\bigg]  \Big\} \, ,
\label{eq:effective-fcnc-explicit}
\end{align} 
with $x_t = m_t^2/m_W^2$. The term $I_{t}(\Lambda,\mu_{w})$ represents the contribution of the RG flow from the scale $\Lambda$ down to $\mu_{w} = m_{t}$ and vanishes if $\Lambda = \mu_{w}$. 

Neglecting the mass of the lighter quark in eq.~\eqref{eq:effective-fcnc-0}, the FCNC Lagrangian may be rewritten as~\cite{Batell:2009jf}
\begin{equation}
    \mathcal{L}_{\text{FCNC}} = ia \sum_{i,j}C_{ij}\bar{q}_{i}(1+\gamma_{5})q_{j}+\text{h.c.},
    \label{eq:effective-fcnc}
\end{equation}
where $C_{ij} \equiv c_{ij}m_{q_{j}}/2f$, where $m_{q_{j}}$ is the mass of the heavier quark among the pair $q_{i}q_{j}$.

The FCNC couplings for the transitions $s\to d$ and $b\to s$ have a huge impact on the ALP phenomenology as they may dominate the production of the ALPs depending on the amounts of kaons and $B$ mesons produced in the given experiment. The value of this coupling is very sensitive to $\Lambda$, growing by two orders of magnitude if increasing the scale from $\Lambda = \mu_{w}$ to $\Lambda \sim 10\text{ TeV}$ (see Fig.~\ref{fig:RG-flow}, right panel).\footnote{As a cross-check of the implementation, we reproduce the values of $c_{ij}$ for the scale $\Lambda = 4\pi \text{ TeV}$ reported in~\cite{Bauer:2021mvw}.}

Taking this into account, we consider two representative choices of $\Lambda$: the one with $\Lambda \equiv \mu_{w}$, and another one with $\Lambda \equiv 1\text{ TeV}$, which is the reference scale used for the PBC BC10 benchmark~\cite{Beacham:2019nyx}.

\section{ALP production}
\label{sec:production}
The ALPs in eq.~\eqref{eq:model-fermions} may be produced by decays of kaons and $B$ mesons, via the mixing with light pseudoscalar mesons $m^{0}$, or by deep inelastic scatterings (DIS).

We describe these production channels in detail in the section below. The amounts of the produced mesons and the DIS cross-section are collision energy dependent. Therefore, to make an experiment-independent comparison, we will consider three collision energies $\sqrt{s}\approx 16\text{ GeV}$, $28\text{ GeV}$, and $13$ TeV (corresponding to the collision energies at DUNE, the SPS beam and the LHC). We took the meson production fractions from the \texttt{SensCalc} repository~\cite{Ovchynnikov:2023cry}.

\subsection{Decays of $B,K$ mesons}

\begin{table}[t]
    \centering   
    \begin{tabular}{c c c c}
       \hline  \vspace{-3mm}  \\ Scale $\Lambda$ & $|C_{bs}|$ & $|C_{bd}|$ & $|C_{sd}|$  \vspace{0.5mm} \\ \hline \vspace{-3mm}  \\
       $m_{t}$  & $2.9\cdot 10^{-11}$ & $5.6\cdot 10^{-12}$ & $4.1\cdot 10^{-15}$ \\
        $1 \text{ TeV}$ & $1.8\cdot 10^{-9}$ & $3.4\cdot 10^{-10}$ & $2.8\cdot 10^{-13}$  \\ \hline 
    \end{tabular}
    \caption{The values of the FCNC couplings from the Lagrangian~\eqref{eq:effective-fcnc} assuming the model~\eqref{eq:model-fermions} with $f = 1\text{ PeV}$ and two scales at which it is defined: $\Lambda = m_{t}$, and $\Lambda = 1\text{ TeV}$.}
    \label{tab:fcnc-couplings}
\end{table}

We will consider the interactions $abs$, $abd$, and $asd$, for which the quark running in the loop is the top quark; the other interactions are heavily suppressed by the Yukawas of lighter quarks and/or CKM elements and are irrelevant. The corresponding decay processes are $B\to a+X_{s}$, $B\to a+\pi$, and $K\to a+\pi$. As we see from eq.~\eqref{eq:effective-fcnc-explicit}, the values of the couplings describing these transitions differ only by the CKM products $V_{ti}^{*}V_{tj}$. This product is the largest for the $b\to s$ transition; however, the other processes are also important. Namely, the $abd$ coupling is suppressed by $|V_{t\to d}/V_{t\to s}|\approx 0.2$; however, the process $B\to a+\pi$ is the only possible above the threshold $B\to K+a$. The relative suppression of the $sd$ coupling is even larger, $|V_{t\to d}|\approx 8\cdot 10^{-3}$. However, depending on the experiment, the number of kaons may be much larger than that of $B$ mesons, which may compensate for this suppression. The values of the corresponding couplings for the two different scales $\Lambda = 1 \text{ TeV}$ or $\Lambda = \mu_{w}$ are given in Table~\ref{tab:fcnc-couplings}. In particular, the value $C_{bs}(1\text{ TeV})$ matches with the one used for the BC10 model~\cite{Beacham:2019nyx,Batell:2009jf}.

Having the operator of the FCNC interaction~\eqref{eq:effective-fcnc}, one may calculate the matrix elements of the processes $B/K \to a + X$, where $X$ is a hadronic state containing an $s$ quark or a $d$ quark. They have the form
\begin{equation}
    \mathcal{M}_{m\to a+m'} = iC_{QQ'}\left(\mathcal{M}^{(S)}_{m\to m'}+\mathcal{M}^{(P)}_{m\to m'}\right),
    \label{eq:FCNC-matrix-element}
\end{equation}
where the parity-even and parity-odd transition matrix elements are
\begin{align}
    \mathcal{M}_{m\to m'}^{(S)} &\equiv \langle m'| \bar{Q}'Q|m\rangle, \nonumber \\ \mathcal{M}^{(P)}_{m\to m'} 
     &\equiv \langle m'| \bar{Q}'\gamma_{5}Q|m\rangle 
     \label{eq:M-meson-transition}
\end{align}
Because of the parity conservation in QCD, if $m,m'$ have the same parity, only $\mathcal{M}_{m\to m'}^{(S)}$ contributes, while for $m'$ having a different parity than $m$ only the $\mathcal{M}^{(P)}_{m\to m'}$ is non-zero. 

The matrix elements~\eqref{eq:M-meson-transition} match with the matrix elements $M_{XX'}$ from eq.~(B.7) from~\cite{Boiarska:2019jym}, used to compute the production of the Higgs-like scalars, which is caused by the similarity of the FCNC operator for ALPs and Higgs-like scalars~\cite{Grzadkowski:1983yp,Leutwyler:1989xj,Haber:1987ua,Chivukula:1988gp}. Therefore, instead of computing the branching ratios using eq.~\eqref{eq:FCNC-matrix-element} one may use the results of Ref.~\cite{Boiarska:2019jym} after the rescaling of the branching ratio with the proper coupling. 

Ref.~\cite{Boiarska:2019jym} used the matrix elements computed using light-cone QCD sum rules and considered the mesons $X_{s} = K, K^{*}(892)$, $K^{*}(1410)$, $K^{*}(1680)$, $K^{0*}(700)$, $K^{0*}(1430)$, $K_{1}(1270)$, $K_{1}(1430), K_{2}^{*}$, and $X_{d} = \pi$.

\begin{figure}[t]
    \centering
\includegraphics[width=0.45\textwidth]{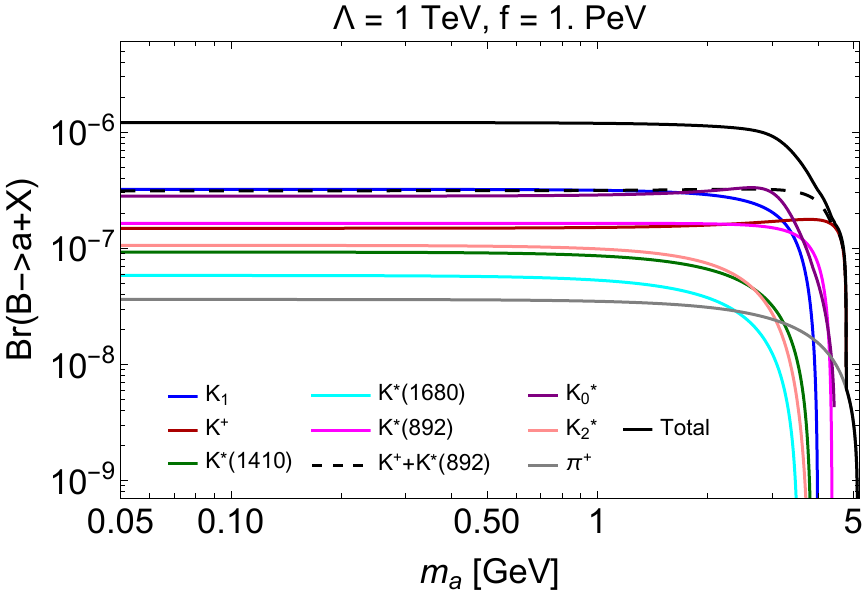}
    \caption{Branching ratios of the 2-body decays $B^{+} \rightarrow X+a$, where $X$ is a hadron that contains an \textit{s} or \textit{d} quark. By the $K^{*}_0$ channel, we denote the final states $K^{*}_{0}(700)$ and $K^{*}_0(1430)$, by $K_1$ --- $K_1(1270), K_1(1400)$. The dashed black lines correspond to the probability of the process $B\to K/K^{*}(892) + a$ considered previously in the literature.}
    \label{fig:br-ratios-B}
\end{figure}

The branching ratios of various decays are shown in Fig.~\ref{fig:br-ratios-B}. Compared to the literature where only the decays $B\to a+ K/K^{*}(892)$ have been considered~\cite{Aloni:2018vki,Beacham:2019nyx,Jerhot:2022chi}, we find almost 4 times larger total production probability. In particular, in the domain of masses $m_{a}\lesssim 1\text{ GeV}$, the dominant production channel is into $K_{1}$ and $K^{0*}$ mesons. 

\hspace{1cm}

\subsection{Mixing with neutral mesons}
\subsubsection{Interaction}
If the ALP is light ($m_{a}\lesssim 2\text{ GeV}$), the description of its hadronic interactions in terms of the $qq$ and $GG$ operators from eq.~\eqref{eq:model-gauge} becomes inadequate since the QCD enters the non-perturbative regime. Instead, light mesons and their interactions represent the strongly interacting sector.  

We follow the existing studies~\cite{Bauer:2020jbp,Bauer:2021mvw,Aloni:2018vki} and obtain the Lagrangian of the ALP interactions with the pseudoscalar mesons $P = \pi,\eta,K,\dots$ by using the matching of the operator~\eqref{eq:model-fermions} with the ChPT Lagrangian. 

Details are provided in Appendix~\ref{app:chpt}; we summarize the main features below. In general, the interaction~\eqref{eq:model-fermions} leads to the kinetic mixings of the ALP with neutral pseudoscalar mesons $m^{0}$. This contrasts with the case of the ALPs coupled to gluons, where the gluon operator also induces the mass mixings~\cite{Aloni:2018vki}. We need to diagonalize the kinetic term to find the relevant interactions. The fields $m^{0}$ entering the Lagrangian are related to the mass eigenstates $m^{0}_{\text{phys}}, a_{\text{phys}}$ by
\begin{equation}
m^{0} \approx m^{0}_{\text{phys}}+ \theta_{m^{0}a}a_{\text{phys}} + \sum_{m^{0'}\neq m^{0}}\theta_{m^{0}m^{0'}}m^{0'}_{\text{phys}}
\label{eq:mass-eigenstates}
\end{equation}
Here, the second term is due to the kinetic mixing with the ALPs, and the third one appears from the mass mixing between the mesons emerging from the minimal ChPT breaking term.

In the limit $m_{s}\gg m_{u,d}$, the mixing angles are 
\begin{widetext}
 \begin{align}
\theta_{\pi^{0}a} &\approx \frac{f_{\pi}F(m_{a})}{f}\frac{m_{a}^{2}}{(m_{a}^{2}-m_{\pi^{0}}^{2})}\left((c_{d}-c_{u})+\delta\frac{m_{\pi^{0}}^{2}}{3}\left[\frac{m_{a}^{2}(c_{d}+2c_{s}+c_{u})}{m_{a}^{2}-m_{\eta'}^{2}}+\frac{2m_{a}^{2}(-c_{s}+c_{u}+c_{d})}{m_{a}^{2}-m_{\eta}^{2}}\right]  \right), \nonumber \\
\theta_{\eta a} &\approx \frac{f_{\pi}F(m_{a})}{f}\sqrt{\frac{2}{3}}\frac{m_{a}^{2}}{m_{a}^{2}-m_{\eta}^{2}}\left((c_{u}+c_{d}-c_{s})-\delta\frac{m_{\pi^{0}}^{2}(c_{u}-c_{d})}{m_{a}^{2}-m_{\pi^{0}}^{2}}\right), \label{eq:mixing-angles} \\ 
\theta_{\eta'a} &\approx \frac{f_{\pi}F(m_{a})}{f}\frac{1}{\sqrt{3}}\frac{m_{a}^{2}}{m_{a}^{2}-m_{\eta'}^{2}}\left(-(c_{d}+2c_{s}+c_{u})-\delta \frac{m_{\pi^{0}}^{2} (c_{u}-c_{d})}{m_{\pi^{0}}^{2}-m_{a}^{2}}\right), \nonumber
 \end{align}
\end{widetext}
where $\delta = (m_{d}-m_{u})/(m_{d}+m_{u})$ is the isospin symmetry breaking parameter, $f_{\pi} \approx 93\text{ MeV}$ is the pion decay constant, and $F(m_{a})$ is a phenomenological function ensuring the drop of the VMD contribution according to quark counting sum rules~\cite{Aloni:2018vki,Jerhot:2022chi}. Similar to~\cite{Aloni:2018vki,Cheng:2021kjg}, in eq.~\eqref{eq:mixing-angles}, we fix $\thet =\arcsin(-1/3)$, motivated by the fact that various phenomenological studies of the effective Lagrangian of the decays of light mesons consider this value. The expressions~\eqref{eq:mixing-angles} are given in terms of the couplings $c_{u},c_{d},c_{s}$ instead of a single $c_{q}$ to account for the RG flow (remind Sec.~\ref{sec:model-details}).

In the first order on the parameter $f_{\pi}/f\ll 1$ and far from resonance domains $m_{a} = m_{m^{0}}$, the ALP and meson masses are left unchanged. Therefore, the only impact of the diagonalization~\eqref{eq:mass-eigenstates} is the appearance of new interactions between the ALPs and mesons. 

The ALP mixing with $\pi^{0}$ emerges either from the RG flow (the first term in eq.~\eqref{eq:mixing-angles}), or via the mixing of unphysical $\pi^{0}$ with $\eta$ and $\eta'$ (the second term).

\begin{figure}[b]
    \centering
    \includegraphics[width=0.45\textwidth]{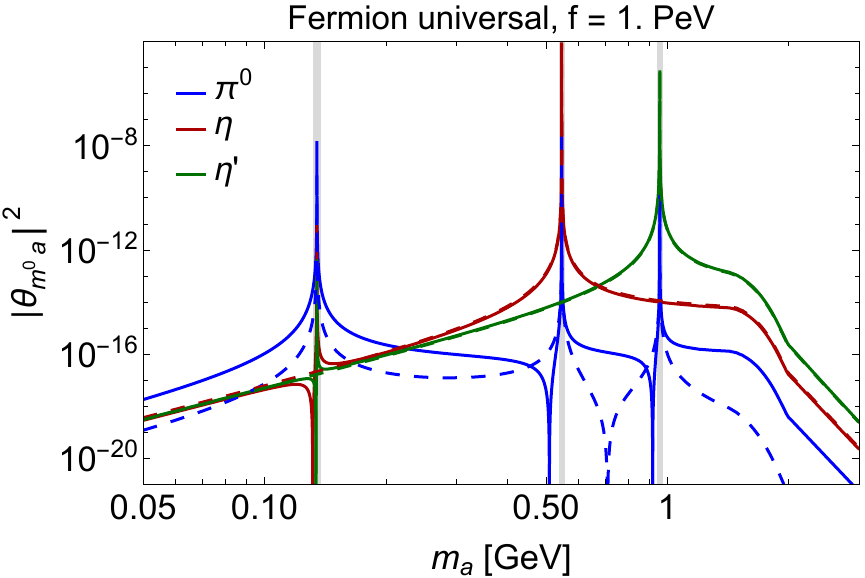}
    \caption{The ALP mass dependence of the square of the mixing angle of the ALP with neutral mesons $\pi^{0}/\eta/\eta'$ (eq.~\eqref{eq:mixing-angles}). The solid lines show the results assuming the scale $\Lambda = 1\text{ TeV}$, while the dashed line corresponds to $\Lambda = m_{t}$. The vertical gray bands correspond to the vicinity of $m_{a} = m_{m^{0}}$ where the approximate description of the ALP interactions via the small mixing with the neutral mesons breaks down (see text for details).}
    \label{fig:mixing-angles}
\end{figure}

The behavior of the squared mixing angles as a function of the ALP mass for the two reference scales $\Lambda = m_{t}$ and $\Lambda = 1\text{ TeV}$ is shown in Fig.~\ref{fig:mixing-angles}.

\subsubsection{Production}
All relevant production processes with ALPs and light neutral mesons occur through their mixings~\eqref{eq:mixing-angles}. We assume that the ALP production cross-section due to the mixing is given by 
\begin{equation}
\sigma_{\text{prod,mixing}} = \sum_{m^{0}=\pi^{0}/\eta/\eta'}\sigma_{\text{prod},m^{0}}\times |\theta_{m^{0}a}|^{2}
\label{eq:production-mixing}
\end{equation}
However, depending on the ALP mass, its kinematics would be different from the corresponding meson kinematics. We follow the procedure described in~\cite{Jerhot:2022chi} to account for this.

We should stress that this description does not account for the ALP mass dependence of the production cross-section. Clearly, the production of the ALPs heavier than $m^{0}$ at the unit mixing angle must be kinematically suppressed, which is not considered in eq.~\eqref{eq:production-mixing}. This point is to be improved in future works. 

\subsection{Deep inelastic scattering process}
\label{sec:production-Drell-Yan}
Another important process of ALP production is deep inelastic scattering. At the parton level and at the leading order in $\alpha_{s}$, it is described by the fusion\footnote{It is important to note that the higher-order processes which we do not consider in this paper may contribute to the flux of the ALPs flying off-axis.}
\begin{equation}
    q + \bar{q} \to a, \quad G + G \to a,
    \label{eq:ALP-production-DIS}
\end{equation}
where for the second process, the matrix element is given by eq.~\eqref{eq:effective-gluon}. The parton model applicability breaks down if the characteristic scale of the process $\sqrt{s_{qq}} = m_{a}$ becomes comparable with $\Lambda_{\text{QCD}}$. We ``turn on'' the process~\eqref{eq:ALP-production-DIS} at $m_{a}= 1.5\text{ GeV}$. 

The DIS process is the only relevant production channel for heavy ALPs with $m_{a}>m_{B}-m_{K}$, given that its kinematic threshold is extended until the center-of-mass energy at the experiment. 

To estimate the cross-section of the process~\eqref{eq:ALP-production-DIS}, we implement the fermionic and gluonic matrix elements in \texttt{MadGraph5}~\cite{Alwall:2014hca} using \texttt{FeynRules}~\cite{Alloul:2013bka,Christensen:2008py} and generate the leading-order processes of the gluon and quark fusion. For the parton distribution function, we use \texttt{NNPDF 3.1 NNLO} set, which is a common choice for FIP sensitivity studies~\cite{Berlin:2018jbm}.  

We have found that the quark fusion is strongly suppressed compared to the gluon fusion. The reason for this is a large value $c_{G}^{\text{eff}}\approx c_{u}+c_{d}+c_{s} \approx 3$ and the fact that the gluonic squared matrix element is proportional to $m_{a}^{3}$ rather than to $m_{a}m_{q}^{2}$ as in the case of the quark fusion, where $m_{q}\ll m_{a}$.

\begin{figure}[t]
    \centering
    \includegraphics[width=0.45\textwidth]{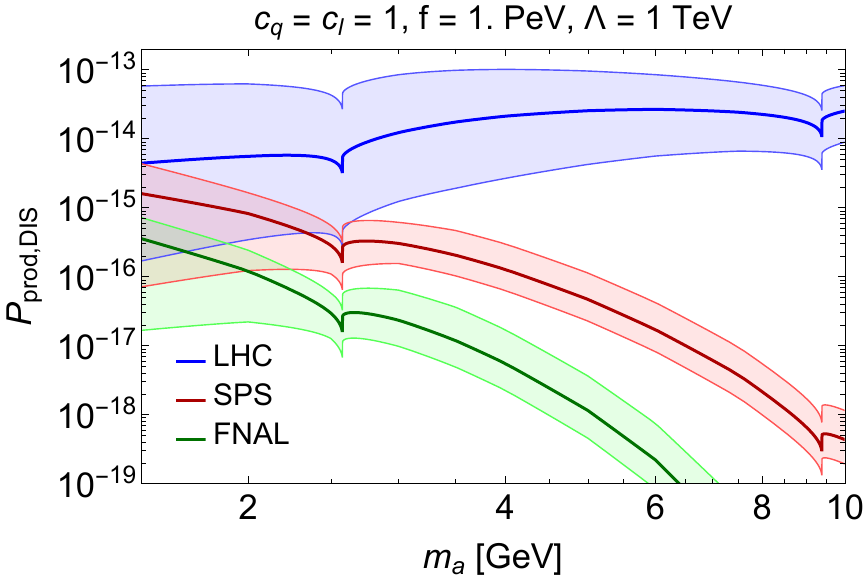}
    \caption{The probability of producing ALPs~\eqref{eq:model-fermions} in the DIS process~\eqref{eq:ALP-production-DIS} at various facilities. The bands show systematics uncertainties, which we obtained by varying the factorization and renormalization scales $\mu_{r}^{2}=\mu_{f}^{2}=m_{a}^{2}$ by a factor of two (see text for details).}
    \label{fig:gluon-fusion-cross-section}
\end{figure}

The cross-section of the gluon fusion at various center-of-mass collision energies is shown in Fig.~\ref{fig:gluon-fusion-cross-section}. It suffers from a large systematical uncertainty, which we illustrate by varying the renormalization and factorization scales $\mu_{r} = \mu_{f} = m_{a}$ by a factor of 2. A huge fraction of this uncertainty comes from the scaling $\sigma_{gg \to a}\propto \alpha_{s}^{2}$, which varies rapidly for small scales $\mu = \mathcal{O}(1\text{ GeV})$. Another important question relevant for the LHC is that the production of light ALPs sits at a very small energy fraction $x = m_{a}^{2}/s_{\text{pp}}$, where the PDFs have a large uncertainty. Including the uncertainties from PDFs would further increase the overall error: by considering several other choices of PDFs, we have found that the cross-section differs by an additional $\mathcal{O}(50\%)$.

Unlike the production via mixing and decays of $B$ mesons, the DIS cross-section is practically independent of the RG flow. Indeed, the effective gluon coupling includes the summation over $u,d,s$ quarks. For the ALP masses of interest, it is proportional to the combination $(c_{u}+c_{d}+c_{s})^{2}$, which for the wide range of the choices of the scale $\Lambda$ changes insignificantly.

\subsection{Discussion}

To compare the contribution of the particular production channels to the ALP yield, we consider the production probabilities per proton collision:
\begin{equation}
    P_{\text{prod}} = \begin{cases} \chi_{m^{0}}\times |\theta_{am^{0}}|^{2}, & \text{mixing} \, , \\ \chi_{B}\times \text{Br}(B\to X_{s}a), & \text{B decays} \, , \\ \frac{\sigma_{\text{DIS}}}{\sigma_{pp}}, & \text{DIS} \; . \end{cases}
    \label{eq:production-probabilities}
\end{equation}
Here, $\chi_{X}$ is the fraction of the produced $X$ particles per proton collision (for $B$, we take both mesons and anti-mesons),  and $\sigma_{pp}$ is the total proton collision cross-section. The mixing angles $\theta_{m^{0}a}$ are taken from eqs.~\eqref{eq:mixing-angles}, and for the branching ratio $\text{Br}(B\to X_{s}a)$ we used Fig.~\ref{fig:br-ratios-B} and Table~\ref{tab:fcnc-couplings}. 

We will not consider here the ALP production by decays of kaons, since the latter are long-lived, and the ALP flux may be heavily affected by the interaction of the kaons with the infrastructure surrounding the kaon production point. The most important factor is their absorption by the material: the absorption length of the kaons is typically smaller than their decay length $3.7E_{K}/m_{K} \text{ m}$. For the beam dump experiments with thick targets, kaons would already be heavily absorbed inside the target. For the LHC-based experiments, the effective kaon decay volume is limited by the detector, and only a small fraction of kaons would decay there due to their huge boosts (see, e.g.,~\cite{Beltran:2023ksw}).\footnote{For the experiments with a thin target like DUNE~\cite{DUNE:2020txw} and NA62~\cite{Hahn:1404985}, whose goal is to maximize the kaon flux within the detector acceptance, kaons bypass the system of magnetic collimators. In such an experiment, one cannot directly obtain the ALP yield based on the number of protons collisions. Estimating sensitivities to FIPs, therefore, requires a dedicated study (see, e.g., \ Ref.~\cite{Ovchynnikov:2022rqj}.} 

\begin{figure}[b]
    \centering
    \includegraphics[width=0.4\textwidth]{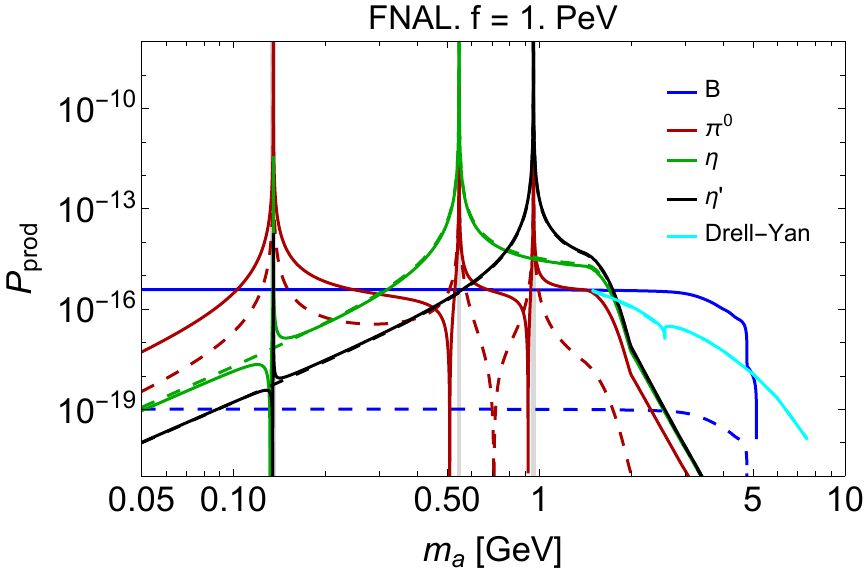} 
    \includegraphics[width=0.4\textwidth]{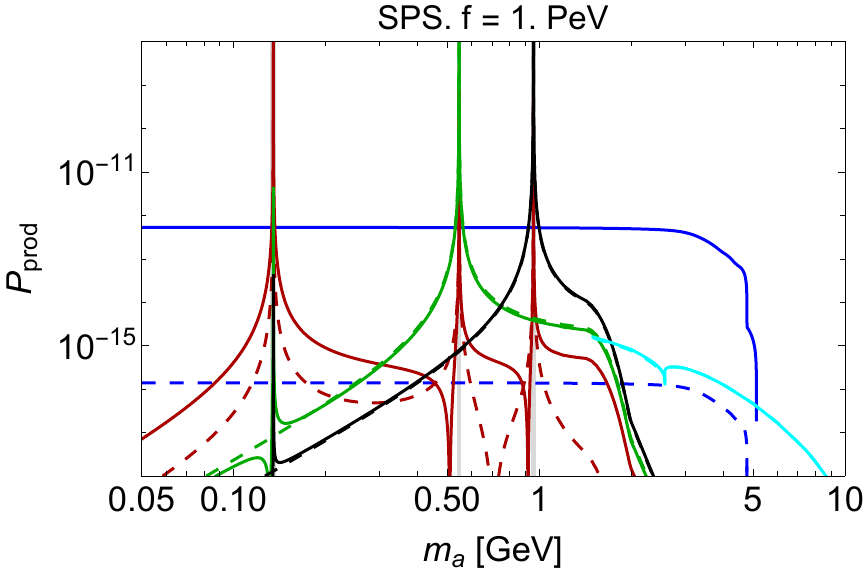} 
    \includegraphics[width=0.4\textwidth]{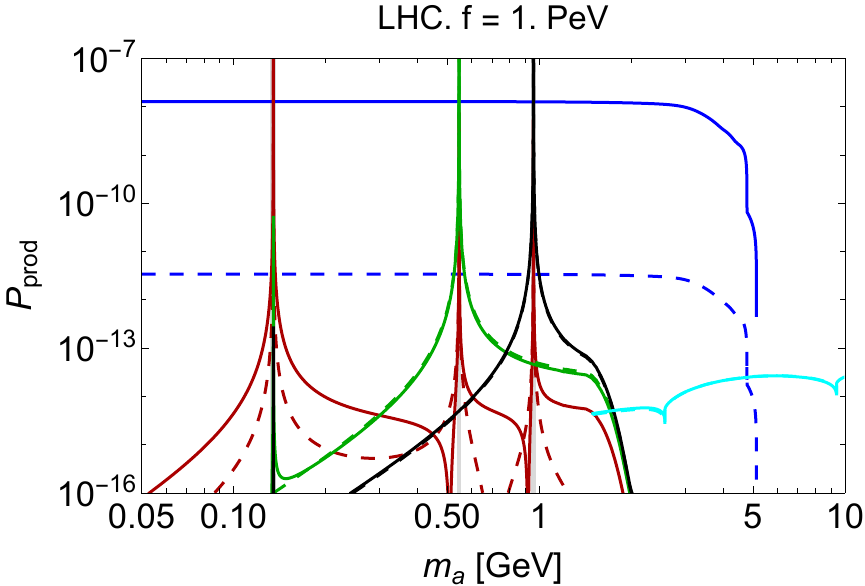} 
    \caption{The ALP production probabilities (eq.~\eqref{eq:production-probabilities}) per proton collision at the LHC, SPS, and FNAL (DUNE), with the collision energies of $\sqrt{s}  =13\text{ TeV}, \approx 28\text{ GeV}$ and $\approx 16\text{ GeV}$ correspondingly, for the model~\eqref{eq:model-fermions}. The probabilities are evaluated for the value of the ALP coupling $f = 1\text{ PeV}$ and assuming two scales $\Lambda$: $\Lambda = m_{t}$ (dashed lines) and $\Lambda = 1\text{ TeV}$ (solid). The meaning of the vertical gray bands is the same as in Fig.~\ref{fig:mixing-angles}. See text for details.}
    \label{fig:alp-production-probabilities}
\end{figure}

The total production probabilities for these energies are shown in Fig.~\ref{fig:alp-production-probabilities}. From the comparison, we see that the dominant production channels at the LHC are decays of $B$ mesons (thanks to the large fraction of the produced $b\bar{b}$ pairs) and Drell-Yan process, independently of the model scale $\Lambda$ chosen. This finding differs strongly from the case of ALPs coupled to gluons~\cite{Chakraborty:2021wda,Bauer:2020jbp}, for which the main production is via the mixing with neutral mesons. The reason may be understood in the following way. All the production channels except for the flavor-violating meson decays receive similar contributions from the tree-level couplings to fermions and gluons $c_{f}, c_{G}$, while the rates of the FCNC decays are very different. Namely, the main contribution to the FCNC coupling is made via the ALP coupling to the top quark, which is absent at the tree level for ALPs coupled to gluons. Instead, it is generated by loops involving gluons~\cite{Chakraborty:2021wda}. The ratio between $bs$ couplings in cases of the fermionic and gluonic ALPs is of the order of $f_{G}/f (\alpha_{s}/\pi)^{2}\ln(\Lambda/m_{t})$, which is $\gg 1$ if assuming similar couplings $f$ and $f_{G} \equiv F/C_{G}$.

The mixing with light mesons may become relevant for the experiments operating at lower energies. Depending on the scale $\Lambda$, it may dominate the total production at SPS at masses $m_{a}\lesssim 2\text{ GeV}$. It also dominates the production at FNAL, even at large $\Lambda$, given the small center-of-mass energy and the correspondingly tiny fraction of produced $B$ mesons. 

\begin{figure}[t!]
    \centering
    \includegraphics[width=0.45\textwidth]{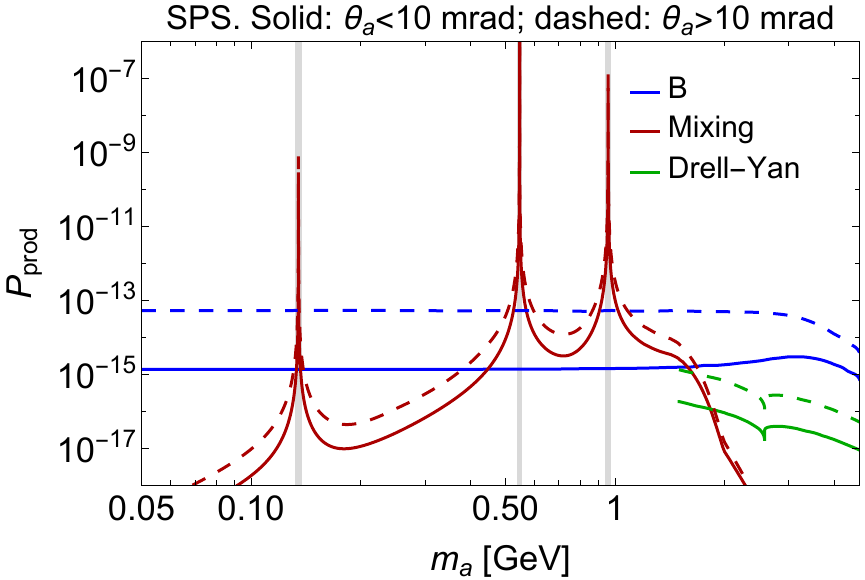}
    \caption{The production probabilities of the ALPs flying in the polar range $\theta<10\text{ mrad}$ (solid lines) and $\theta>10\text{ mrad}$ (dashed lines) at SPS, assuming the scale $\Lambda = 300\text{ GeV}$. By the curves ``mixing'', we summarize the production of the ALPs via the mixing angles with the mesons $\pi^{0},\eta,$ and $\eta'$. The value $f = 1\text{ PeV}$ is assumed.}
    \label{fig:alp-production-probabilities-angular}
\end{figure}

To conclude this discussion, we emphasize that the hierarchy of the production channels may change depending on the placement of the experiment with respect to the proton beam axis. Generically, the angular distributions of the light ALPs from $B$ decays and those produced by the Drell-Yan process are broader than the distribution from the mixing with light mesons. Therefore, if these production channels provide similar overall amounts of ALPs, the mixing with the mesons would dominate for on-axis experiments with small angular coverage, while the other channels dominate for off-axis experiments. To demonstrate this point, in Fig.~\ref{fig:alp-production-probabilities-angular}, we show the production probabilities for the ALPs flying within the polar range $\theta < 10\text{ mrad}$ and $\theta > 10\text{ mrad}$ at SPS, assuming $\Lambda = 300\text{ GeV}$.

\section{Decay modes}
\label{sec:decays}
\subsection{Decays into leptons and photons}

The matrix element of the decay of ALPs into a pair of photons is given by
\begin{equation}
    \mathcal{M}_{a\to \gamma\gamma} = \frac{1}{f}\frac{\alpha_{\text{EM}}}{\pi}c^{\text{eff}}_{\gamma} F_{\mu\nu}(p_{\gamma 1})\tilde{F}^{\mu\nu}(p_{\gamma 2})
\end{equation}
Depending on the ALP mass, the effective coupling $c_{\gamma}^{\text{eff}}$ is
\begin{equation}
    \cg = c_{\gamma}^{\text{loop,l}}+ \begin{cases}
    c_{\gamma}^{\text{VMD}}, \quad m_{a}<m_{a}' \\ c_{\gamma}^{\text{loop,q}}, \quad m_{a}\geq m_{a}'
    \end{cases}
\end{equation} 
where $m'_a \simeq 2$~GeV is similar to the matching scale between the ALP's ChPT and QCD perturbative decays (see the next subsection).

The VMD contribution $c_{\gamma}^{\text{VMD}}$ originates from the mixing of the vector mesons $\rho,\omega,\phi$ with photons. The corresponding contributions are
\begin{multline}
    c_{\gamma}^{\text{VMD}} = -\frac{1}{9}F(m_{a})\big(4\sqrt{6}\theta_{\eta a}+7\sqrt{3}\theta_{\eta'a}+9\theta_{\pi^{0}a}\big)
\end{multline}
We have checked that this coupling approximately reproduces the widths of the anomalous decays $m^{0}\to 2\gamma$ with $m^{0} = \pi^{0}/\eta/\eta'$ in the symbolic limit  $\theta_{m^{0'}a} = \delta_{m^{0}m^{0'}}$, $f = f_{\pi}$, and $m_{a} = m_{m^{0}}$.

The loop contribution is given by triangle diagrams with fermions $f = l,q$ running inside the loop~\cite{Bauer:2017ris,Bauer:2021mvw}:
\begin{equation}
    c_{\gamma}^{\text{loop,f}} = \sum_{f}2N_{c}^{f}Q_{f}^{2}c_{f}B_{1}(\tau_{f}),
\end{equation}
where $N_{c}^{q} = 3$, $N_{c}^{l} = 1$, $Q_{f}$ is the charge of the fermion, and $B_{1}$ is from eq.~\eqref{eq:B1}. 

Decay widths into lepton pairs $\ell^{+}\ell^{-} = e^{+}e^{-}$, $\mu^{+}\mu^{-}$, $\tau^{+}\tau^{-}$ are described by the formula
\begin{equation}
     \Gamma(a\to \ell^{+}\ell^{-}) = \frac{c_{\ell}^{2}m_{a}m_{\ell}^{2}\sqrt{1-\frac{4m_{\ell}^{2}}{m_{a}^{2}}}}{2\pi f^{2}},
\end{equation}

The total width into leptons and the width into photons are shown in Fig.~\ref{fig:hadronic-widths}.

\subsection{Hadronic decays}

Let us now discuss the hadronic decays of ALPs. For $m_{a} \gg\Lambda_{\text{QCD}}$, it is adequate to describe these decays by decays into quarks and gluons~\cite{Bauer:2021mvw}:
\begin{align}
    \Gamma(a\to q\bar{q}) &= \frac{N_{c}c_{q}^{2}m_{a}m_{q}^{2}\sqrt{1-\frac{4m_{X_{q}}^{2}}{m_{a}^{2}}}}{2\pi f^{2}}, \\ \Gamma(a\to GG) &= |c_{G}^{\text{eff}}|^{2}\frac{\alpha_{s}^{2}\left(1 + \frac{83\alpha_{s}}{4\pi} \right)m_{a}^{3}}{8\pi^{3} f^{2}},
\end{align}
where $c_{G}^{\text{eff}}$ is given by eq.~\eqref{eq:cGeff}, and $N_{c} = 3$ is the number of colors. To approximately account for hadronization, when considering the kinematics, we replace the quark's mass with the mass of the lightest meson containing the given quark~\cite{Gunion:1989we,Winkler:2018qyg}. For instance, for decays into $c\bar{c}$, $X_{q}$ is a $D$ meson.

Because of the same reason as it was discussed in the context of the Drell-Yan production (Sec.~\ref{sec:production-Drell-Yan}), decays into gluons dominate over decays into light quarks $u\bar{u}, d\bar{d}, s\bar{s}$. However, above the $D$-meson pair production threshold, the decay into $c\bar{c}$ contributes a sizable fraction of the ALP total width because of a large $c$ mass. 

For $m_{a}=\mathcal{O}(1\text{ GeV})$, perturbative QCD breaks down, and one should use ChPT. To describe the palette of various decays (e.g., $\eta/\eta' \to \pi \pi \gamma, \eta'\to 4\pi$, or $\eta'\to \eta \pi \pi$) in agreement with the experimental data, the minimal ChPT is supplemented by phenomenological Lagrangians of the interactions of the pseudoscalar mesons $P$ with vector~\cite{Fujiwara:1984mp,Guo:2011ir}, scalar~\cite{Fariborz:1999gr}, and tensor mesons~\cite{Guo:2011ir}, with the operators and their couplings being fixed by theoretical arguments (such as the chiral symmetry or anomaly matching conditions) and to match the experimental data on interactions of $P$. These mesons may contribute to the matrix elements as intermediate states; one example is the mixing of neutral vector mesons $\rho^{0},\omega,\phi$ with photons. The ChPT width should match the parton-level width at some mass $m_{a}\simeq 1\text{ GeV}$.

Ref.~\cite{Aloni:2018vki} followed this data-driven approach to describe the decays of the ALPs coupled to gluons; however, their approach also applies to the ALPs coupled to fermions. Ref.~\cite{Cheng:2021kjg} repeated the analysis of~\cite{Aloni:2018vki} with some modifications for the ALPs with a non-universal explicitly isospin-breaking coupling to quarks. 
\begin{figure*}
    \centering
    \includegraphics[width=0.65\textwidth]{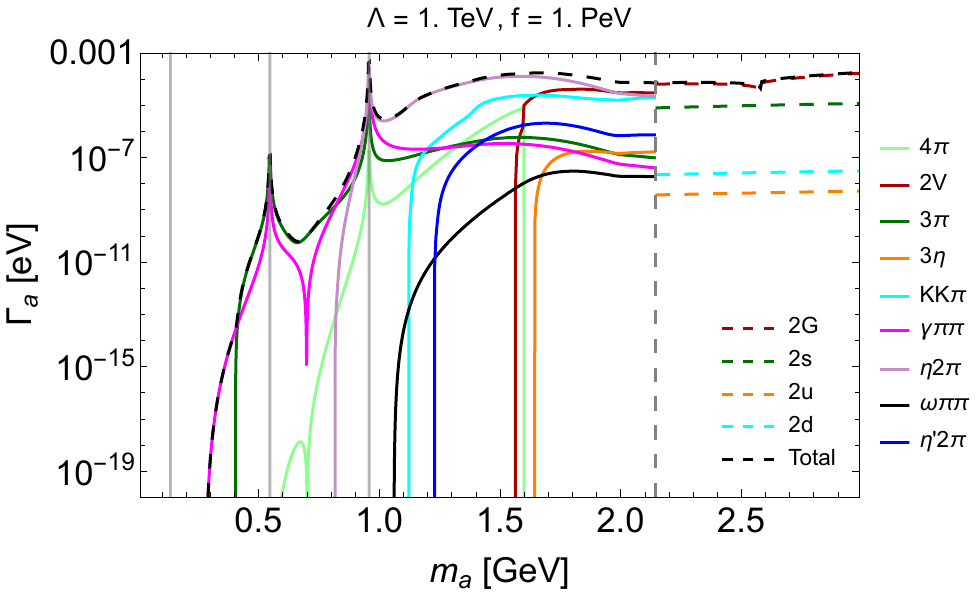} \\  \includegraphics[width=0.65\textwidth]{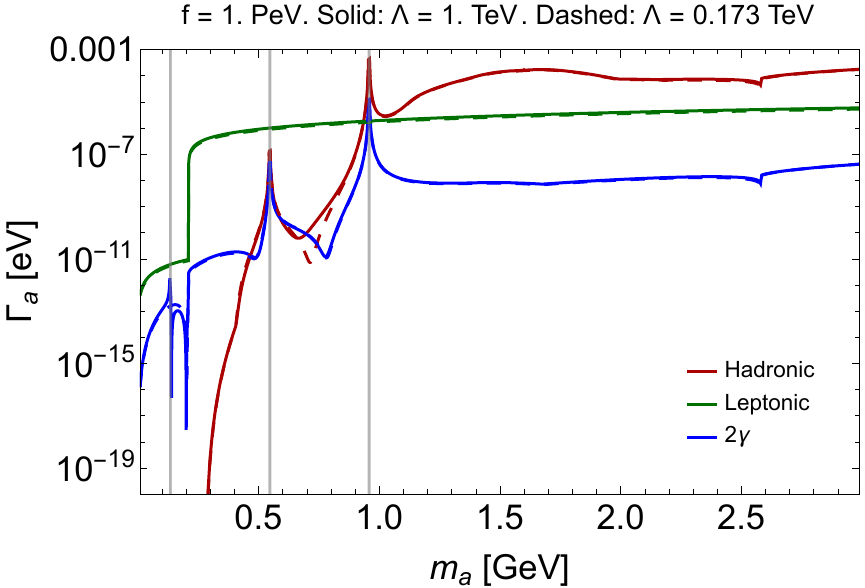}
    \caption{Summary of decays of the ALPs with the universal coupling to fermions~\eqref{eq:model-fermions}. Top panel: hadronic decay widths, assuming the scale $\Lambda = 1\text{ TeV}$. For the ALP mass $m_{a}\lesssim 2\text{ GeV}$, ChPT describes decays of ALPs, while at higher masses, they may be approximated by the perturbative QCD; the vertical dashed line shows the matching mass. \textit{Bottom panel}: the total non-hadronic (including di-$e$,$\mu$,$\tau$, and $\gamma$ processes) and hadronic widths. The solid lines correspond to the scale $\Lambda = 1\text{ TeV}$, while the dashed ones correspond to $\Lambda = m_{t}$. The vertical gray bands show the vicinity of $m_{a} = m_{\pi^{0}}/m_{\eta}/m_{\eta'}$ where the description via the ALP-meson mixing (and hence the results for the hadronic widths) breaks down.}
    \label{fig:hadronic-widths}
\end{figure*}

In our analysis, we incorporate the ChPT Lagrangian (following the references above) in the \texttt{Mathematica} notebook accompanying the paper and calculate the matrix elements and decay widths for various processes (see Appendices~\ref{app:mathematica-notebook},~\ref{app:chpt} for details). We include the decay channels $a\to \eta 2\pi,\eta'2\pi,4\pi$, $3\pi,\gamma 2
\pi$, $3\eta$, $\omega 2\pi$, and $2V$, where $V = \omega, K^{*},\phi$. As a cross-check, we reproduce the results of the SM decay widths of the mesons $\eta$ and $\eta'$ in the limit when the ALP matches them, i.e., when $\theta_{m^{0}a} = 1$, $f = f_{\pi}$, and $m_{a} = m_{m^{0}}$. We have also qualitatively reproduced the results from~\cite{Aloni:2018vki} for the model of the ALPs coupled to gluons (see a discussion in Appendix~\ref{app:chpt}). 

The summary of the hadronic widths for the ALPs is shown in Fig.~\ref{fig:hadronic-widths}. The decays of low-mass ALPs $m_{a}\lesssim 1\text{ GeV}$ are saturated by $a\to 3\pi$, $a\to \gamma\pi\pi$. At higher masses, decays into $\eta \pi \pi$, $4\pi$, and $2V$ become the dominant channels. The ChPT width matches with the width in the perturbative regime at $m_{\text{match}}\simeq 2\text{ GeV}$.

\subsection{Discussion}
\begin{figure*}[t]
    \centering
    \includegraphics[width=0.45\textwidth]{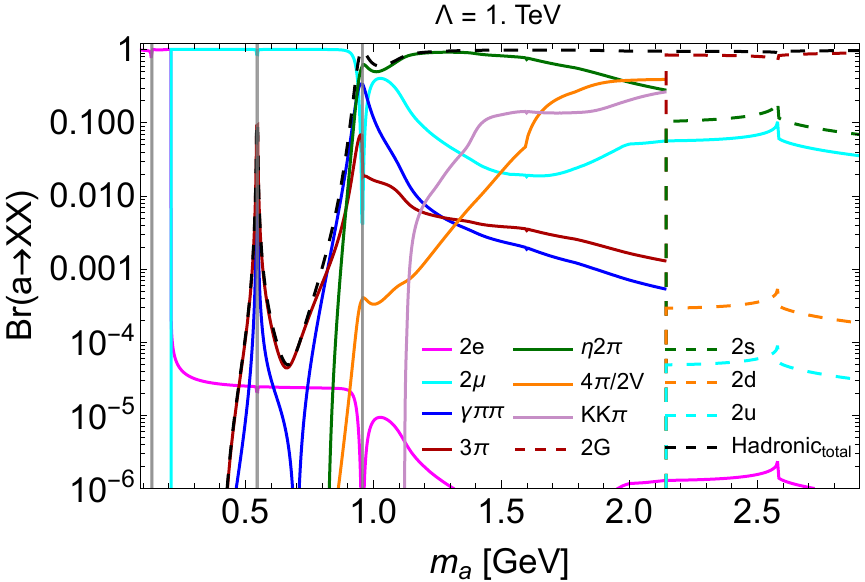}~\includegraphics[width=0.45\textwidth]{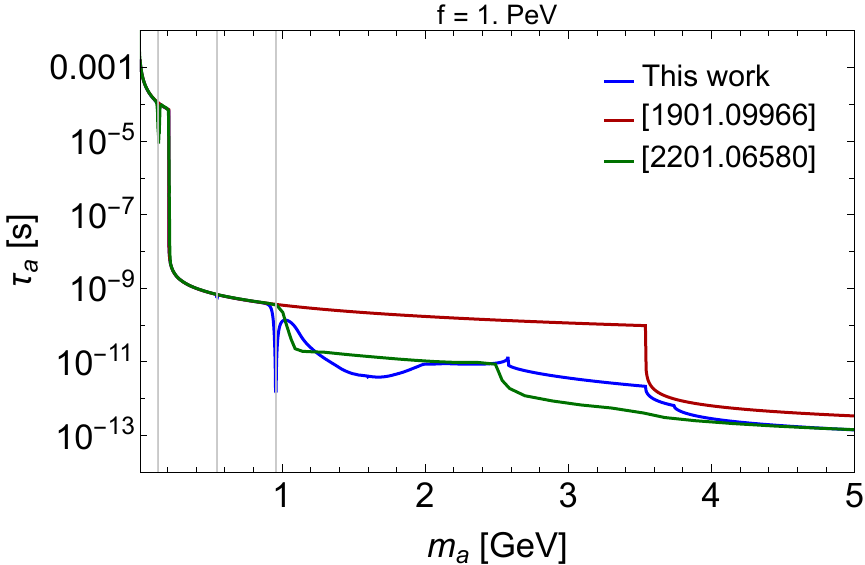}
    \caption{\textit{Left panel}: branching ratios of the ALP decays into important final states (see Fig.~\ref{fig:hadronic-widths} for the description of the states) for the model~\eqref{eq:model-fermions}. \textit{Right panel}: the ALP lifetime assuming $f = 1\text{ PeV}$. The red and blue lines show, correspondingly, the lifetime in the approximation of using only leptonic width as in~\cite{Beacham:2019nyx} and our result (both assuming $\Lambda = 1\text{ TeV}$), while the green line the results from~\cite{Ferber:2022rsf} evaluated for the scale $\Lambda = 4\pi \text{ TeV}$.}
    \label{fig:new-vs-old}
\end{figure*}

The leptonic, photonic, and hadronic widths are compared in Fig.~\ref{fig:hadronic-widths}.  In Fig.~\ref{fig:new-vs-old}, we show the branching ratios of the ALP decays into various final states. From the figures, we see that photonic decays are always sub-dominant, while hadronic widths are irrelevant for the phenomenology of light ALPs with $m_{a}\lesssim 1\text{ GeV}$, where leptonic decays dominate. For heavier ALPs, however, decays into hadrons dominate, increasing the total width by up to a factor of 100. This conclusion is in qualitative agreement with the paper~\cite{Domingo:2016yih}, which studied a somewhat different model of a CP-odd scalar. We emphasize that in the mass range around 1 GeV, the hadronic decay width is significantly larger than the one obtained in Refs.~\cite{Bauer:2020jbp,Ferber:2022rsf} using perturbative QCD.

The decay palette above 1 GeV is also qualitatively similar to the case of the ALPs coupled to gluons. This is because, for both of these models, the ALPs have mixing with the three neutral pseudoscalar mesons $\pi^{0}/\eta/\eta'$.  

Interestingly, the choice of the scale $\Lambda$ practically does not influence the decay phenomenology. This is because it affects only the decays where the mixing with pions dominates among the others. These are the decays into $3\pi$ and $\gamma \pi \pi$, which are important only in the mass range $m_{a}\lesssim 1\text{ GeV}$ where leptonic decay widths are much larger.

To further stress the importance of the hadronic decays and finalize the discussion, we show the mass dependence of the ALP lifetime as computed in this work and the one widely considered in the past~\cite{Beacham:2019nyx}, when only leptonic decays have been included. In the mass range $2m_{\mu}<m_{a}<2m_{\tau}$, the full width in the description from~\cite{Beacham:2019nyx} is saturated by dimuon decay. With hadronic decays being included, the branching ratio $\text{Br}(a\to \mu\mu)$ is $<10\%$ for masses $m_{a}\gtrsim 1\text{ GeV}$. 

In Fig.~\ref{fig:new-vs-old}, we show the branching ratios of the ALP decays into various final states (left panel) and the ALP lifetime (right panel). For the lifetime, we compare the predictions assuming the revised phenomenology and the description from past works: Ref.~\cite{Beacham:2019nyx}, which neglects hadronic decay modes and is widely used by the experiments community to derive constraints and sensitivities, and Ref.~\cite{Ferber:2022rsf}, which, following~\cite{Bauer:2020jbp,Bauer:2021mvw}, approximates the hadronic decays by the decay $a\to 3\pi$ at $m_{a}\lesssim 1\text{ GeV}$ and by the decays into a gluon and quark pairs above this mass. The lifetime from~\cite{Beacham:2019nyx} is always much larger for $m_{a}\gtrsim 1\text{ GeV}$. The lifetime from~\cite{Ferber:2022rsf} coincides with our result for the ranges $m_{a}\lesssim 1\text{ GeV}$, $m_{\text{match}}<m_{a}<2m_{c}$, and $m_{a}\gtrsim 2m_{D}$. The origin of the discrepancy in the range $2m_{c}<m_{a}\lesssim 4\text{ GeV}$ is that Ref.~\cite{Ferber:2022rsf} turns on the decay $a\to c\bar{c}$ above $2m_{c}$, even though this decay is kinematically impossible until the $2D$ threshold.

\begin{figure}[t]
    \centering
    \includegraphics[width=0.45\textwidth]{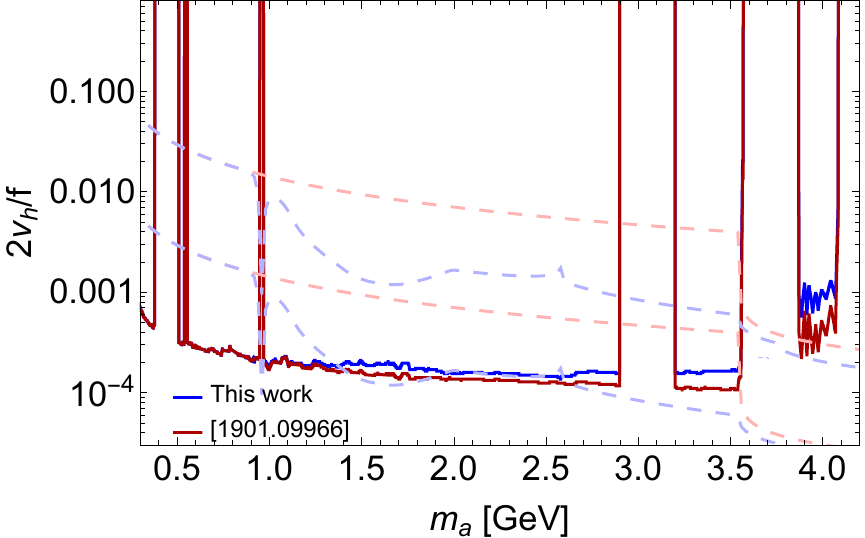}
    \caption{Re-interpretation of the model-independent LHCb constraints from the searches $B\to K+(\text{FIP}\to)\mu\mu$ reported in~\cite{LHCb:2015nkv,LHCb:2016awg} for the model of ALPs with the fermion coupling~\eqref{eq:model-fermions}, assuming $\Lambda = 1\text{ TeV}$, for the plane $m_{a}-2v_{h}/f$, where $v_{h} = 246\text{ GeV}$ is the Higgs VEV. The blue solid line: constraints assuming the ALP phenomenology obtained in this work. The red solid line: if assuming the phenomenology from~\cite{Beacham:2019nyx}, which only includes leptonic decays. The light lines of the same colors show the ALP decay lengths $c\tau_{a}\langle \gamma_{a}\rangle = 1\text{ cm}$ and $1\text{ m}$ assuming the corresponding phenomenology.}
    \label{fig:lhcb-bound}
\end{figure}

The previously neglected production and decay modes are expected to significantly change the landscape of the past constraints and future searches for ALPs. For example, let us consider searches for $B\to K(a\to)\mu\mu$ performed at LHCb~\cite{LHCb:2015nkv,LHCb:2016awg}. It is sensitive to the total ALP decay width as well as to the branching ratio $\text{Br}(a\to \mu\mu)$. The constraints to ALPs are shown in Fig.~\ref{fig:lhcb-bound}, where, for comparison, we display the bound obtained assuming the ALP phenomenology description from Ref.~\cite{Beacham:2019nyx} and the one obtained in this work. The updated constraints are weaker in the domain of large masses $m_{a}\gtrsim 1\text{ GeV}$. Interestingly, for the revised phenomenology, the lower bound of the constraint lies in the regime where ALPs are short-lived and mainly decay within the detector, whereas for the old phenomenology, it mainly belongs to the parameter space of long-lived ALPs. We will revise further existing constraints, consider the ones previously not accounted for in the literature, and derive sensitivities for future experiments in a forthcoming work~\cite{ALP-2}. 

\section{Conclusions}
\label{sec:conclusions}

The model of ALPs universally coupled to fermions is considered by the physics beyond colliders (PBC) group as one of the benchmark models to test the potential of various experiments to explore the parameter space of feebly-interacting particles. Therefore, understanding the phenomenology of such ALPs is an important and timely question. In this work, we have revised the production and decay modes of ALPs at hadronic accelerator experiments, also considering the impact of the renormalization group (RG) flow of their couplings depending on the scale $\Lambda$ at which the model is defined (see Sec.~\ref{sec:model-details} and in particular Fig.~\ref{fig:RG-flow}).

For the production (Sec.~\ref{sec:production}), we have considered decays of kaons and $B$ mesons, the mixing with neutral mesons $\pi^{0}/\eta/\eta'$, and the Drell-Yan process. For the production via mixing, we have found that the RG flow is very important, sizably changing the mixing angle squared between the ALPs and $\pi^{0}$ (Fig.~\ref{fig:mixing-angles}). For the production from $B$ mesons, we have included the decays $B\to X_{s}+ a$, with $X_{s}$ being heavy kaon resonances $K^{*0}, K_{1},K_{2},\dots$, which have not been considered previously in this context in the literature and increase the total production branching ratio by a factor of 3--4 for light ALPs (Fig.~\ref{fig:br-ratios-B}). Our results apply also to generic ALPs, provided that the low-energy Lagrangian describing the decay has the same operator expression. For the Drell-Yan production, we have considered the leading-order fusion processes and shown that the cross-section suffers from a large systematic uncertainty (Fig.~\ref{fig:gluon-fusion-cross-section}). 
 
Depending on the scale $\Lambda$ at which the ALP model~\eqref{eq:model-fermions} is defined and the collision energy, we have found that any of these processes may dominate the production (Fig.~\ref{fig:alp-production-probabilities}). In particular, at DUNE collision energy, the production via mixing is the main production channel of the ALPs with mass below 2--3 GeV, while at larger masses, decays of $B$ mesons are the main channel. At SPS energies, the hierarchy of production channels depends heavily on the scale $\Lambda$. Namely, if $\Lambda$ is close to the EW scale, the main channels are mixing with mesons and the Drell-Yan process. Once $\Lambda$ departs from the EW scale, decays via $B$ mesons dominate. The main production channel may also depend on the geometric placement of the experiment (see Fig.~\ref{fig:alp-production-probabilities-angular}). Finally, at the LHC, $\Lambda$ practically does not influence the hierarchy and affects only the magnitude of the ALP flux.

We have studied the full palette of the ALP decays (Sec.~\ref{sec:decays}), including the hadronic ones that were missing previously. In particular, we have found (Fig.~\ref{fig:hadronic-widths} and also Fig.~\ref{fig:new-vs-old}) that leptonic decays are the main channels for light ALPs with $m_{a}\lesssim 1\text{ GeV}$, while the hadronic decays dominate at higher masses, increasing the total ALP width by up to a factor of 100. Contrary to the production case, the decay widths are only weakly sensitive to the choice of $\Lambda$.

To simplify the use of our results by the community, we have implemented the ALP phenomenology studied in this work in a \texttt{Mathematica} notebook accompanying the paper. We have also implemented the model in \texttt{SensCalc}~\cite{Ovchynnikov:2023cry} -- a public code evaluating sensitivities of different experiments.

\section*{Acknowledgements}

We thank Pilar Coloma, Torben Ferber, Joerg Jaeckel, Jan Jerhot, Felix Kling, Thomas Schwetz, Yotam Soreq and Susanne Westhoff for helpful discussions. This work was partially funded by the Deutsche Forschungsgemeinschaft (DFG) through the Emmy Noether Grant No. KA 4662/1-2 and grant 396021762~--~TRR~257. MO received support from the European Union's Horizon 2020 research and innovation program under the Marie Sklodowska-Curie grant agreement No. 860881-HIDDeN. GG thanks the Doctoral School  ``Karlsruhe School of Elementary and Astroparticle Physics: Science and Technology (KSETA)” for financial support through the GSSP program of the German Academic Exchange Service (DAAD).

\newpage

\bibliographystyle{apsrev.bst}
\bibliography{bib.bib}

\onecolumngrid 

\appendix

\newpage

\section{ChPT with ALPs}
\label{app:chpt}

Let us, for completeness, assume that both the quark couplings $c_{q}$ from eq.~\eqref{eq:model-fermions} and the gluon coupling $c_{G}$ in eq.~\eqref{eq:effective-gluon} are present in the Lagrangian. Both of them may contribute to the ChPT interactions. To this end, let us first convert the gluon coupling in eq.~\eqref{eq:effective-gluon} to the pure quark sector by performing the following chiral rotation of the light quarks $q =(u,d,s)$~\cite{Krauss:1986bq,Bardeen:1986yb,Srednicki:1985xd,Georgi:1986df}:
\begin{equation}
    q \to \mathcal{U}q, \quad \mathcal{U} = \exp\left[-ic_{G}\frac{a}{f}\kappa_{q}\gamma_{5}\right] \; ,
\end{equation}
where $(\kappa_{q})_{ij} = \delta_{ij}m_{q_{i}}^{-1}/(m_{u}+m_{d}+m_{s})$ is fixed in order to prevent any mass mixing between the mesons $\pi^{0}$ and $\eta_{0}$ with $a$. The hadronic part of the Lagrangian becomes 
\begin{equation}
    \langle\mathcal{L}_{\text{model 1,\text{hadr}}}\rangle_{c,b,t} = \bar{q} \, \mathcal{U} m \, \mathcal{U} q +\frac{\partial^{\mu}a}{f} (c_{q}+c_{G}\kappa_{q})\bar{q}\gamma_{\mu}\gamma_{5}q \; ,
    \label{eq:model-1-hadr}
\end{equation} 
where we have neglected the off-diagonal quark coupling $c_{sd}$ generated by integrating out the top quark due to its strong CKM suppression. The relevant ChPT Lagrangian then is~\cite{Bauer:2020jbp,Aloni:2018vki}
\begin{multline}
    \mathcal{L}_{\text{ChPT,min}} =\frac{1}{2}(\partial_{\mu}a)^{2}-\frac{m_{a}^{2}}{2}a^{2}+ \frac{f_{\pi}^{2}}{2}B_{0}\text{Tr}\left[\Sigma \hat{m}^{\dagger}_{q}+\hat{m}_{q}\Sigma^{\dagger}\right]+\frac{f_{\pi}^{2}}{4}\text{Tr}\left[ D_{\mu}\Sigma D^{\mu}
    \Sigma^{\dagger}\right] + \\ \frac{f_{\pi}^{2}}{2}\frac{\partial_{\mu}a}{f}\text{Tr}[(\hat{c}_{q}+c_{G}\kappa_{q})(\Sigma D^{\mu}\Sigma^{\dagger}-\Sigma^{\dagger}D^{\mu}\Sigma)] \; ,
    \label{eq:lagr-chpt-1}
\end{multline}
where $c_{q} = \text{diag}(c_{u},c_{d},c_{s})$,
\begin{equation}
\hat{m}_{q} =  \exp\left[-ic_{G}\frac{a}{f}\kappa_{q}\right]m_{q}\exp\left[-ic_{G}\frac{a}{f}\kappa_{q}\right] \; ,
\end{equation}
$\Sigma$ is the matrix of the pseudoscalar mesons
\begin{equation}
    \Sigma = \exp\left[ \frac{2i\mathcal{P}}{f_{\pi}}\right], \quad \mathcal{P} = \frac{1}{\sqrt{2}}\begin{pmatrix}
        \frac{\pi^{0}}{\sqrt{2}}+\frac{\eta}{\sqrt{3}}+\frac{\eta'}{\sqrt{6}} & 
    \pi^{+}& K^{+} \\ \pi^{-} & -\frac{\pi^{0}}{\sqrt{2}}+\frac{\eta}{\sqrt{3}}+\frac{\eta'}{\sqrt{6}}  & K^{0} \\ K^{-} & \bar{K}^{0} & -\frac{\eta}{\sqrt{3}} +2\frac{\eta'}{\sqrt{6}},
    \end{pmatrix} 
\end{equation}
$D_{\mu}\Sigma = \partial_{\mu}\Sigma + ieA_{\mu}\Sigma$ is the covariant derivative.

We also need to include the phenomenological Lagrangian of the interactions of pseudoscalar mesons with other mesons: anomalous WZW interactions and interactions with vector~\cite{Fujiwara:1984mp,Guo:2011ir}, scalar~\cite{Fariborz:1999gr}, and tensor meson $f_{2}$~\cite{Cheng:2021kjg} (see also~\cite{Guo:2011ir}):
\begin{align}
    \mathcal{L}_{\text{vec+an}} = & -\frac{3g^{2}}{8\pi^{2}f_{\pi}}\epsilon^{\mu\nu\alpha\beta}\text{Tr}[P(x)\partial_{\mu}V_{\nu}(x)\partial_{\alpha}V_{\beta}(x)]+\frac{7}{60\pi^{2}f_{\pi}^{5}}\epsilon^{\mu\nu\alpha\beta}\text{Tr}[P(x)\partial_{\mu}P\partial_{\nu}P\partial_{\alpha}P\partial_{
    \beta}P]\\ & +2f_{\pi}^{2}\text{Tr}\left|gV_{\mu}-eA_{\mu}Q-\frac{i}{2f_{\pi}^{2}}[P,\partial_{\mu}P]\right|^{2}
    \\ & - g_{f_2 \pi \pi} \frac{f_\pi^2}{4}\text{Tr}\left[\left(\partial^\mu \Sigma^{\dagger} \partial^\nu \Sigma-\frac{1}{2} g^{\mu \nu} \partial^\alpha \Sigma^{\dagger} \partial_\alpha \Sigma\right) \boldsymbol{f}_{\mathbf{2}}\right] \phi_{\mu \nu} +\mathcal{L}_{\text{scalar}}
    \label{eq:lagr-chpt-2}
\end{align}
Here, $g \approx m_{\rho}/\sqrt{2}f_{\pi}$, $Q = \text{diag}[2/3,-1/3,-1/3]$ is the quark charge matrix, $V_{\mu}$ is the matrix of vector mesons, 
\begin{equation}
    V_{\mu} = \frac{1}{\sqrt{2}}\begin{pmatrix}
        \frac{\rho^{0}+\omega}{\sqrt{2}} & \rho^{+} & K^{*+} \\ \rho^{-} & \frac{-\rho^{0}+\omega}{\sqrt{2}} & K^{*0} \\ K^{*-} & \bar{K}^{*0} & \phi
    \end{pmatrix} 
\end{equation}
and $A_{\mu}$ is the EM field. Next, $\mathcal{L}_{\text{scalar}}$ is given by eq.~(A1) from~\cite{Fariborz:1999gr}. The tensor meson $f_2$ is denoted by $\phi_{\mu \nu}$, while $\bm{f_{2}}$ is the SU(3) generator of the tensor meson. The coupling $g_{f_2 \pi \pi}=13.1 \text{ GeV}^{-1}$~\cite{Cheng:2021kjg}.

Having the Lagrangians~\eqref{eq:lagr-chpt-1},~\eqref{eq:lagr-chpt-2}, we calculate the various contributions to the matrix elements of the ALP production and decay (see Appendix~\ref{app:mathematica-notebook} for the implementation). 

When calculating the decay matrix elements, we mostly follow the assumptions considered in~\cite{Cheng:2021kjg} based on observational data and unitarity requirements. As an example, we artificially set the contact VMD terms originating from the square of the last summand of the second line of eq.~\eqref{eq:lagr-chpt-2} to zero if $m_{a}>m_{\eta'}$.

\subsection{Comparison with ref.~\cite{Aloni:2018vki}}

There are some differences in the description of the ALP decays from~\cite{Aloni:2018vki} and~\cite{Cheng:2021kjg} (and hence our approach). For instance, unlike~\cite{Cheng:2021kjg}, ref.~\cite{Aloni:2018vki} does not include the contributions of $\mathcal{L}_{\text{vec+an}}$ to the decays $a\to 3\pi$, which changes the corresponding width by orders of magnitude. It is crucial since this width dominates the ALP decays in the mass range $m_{a}\lesssim 1\text{ GeV}$. Another difference is that the vector meson contribution to the widths $a\to KK\pi$ is included in~\cite{Cheng:2021kjg} but not in~\cite{Aloni:2018vki}.

Yet another difference is in the interaction sector with scalar mesons. We have used the interaction Lagrangian directly from Appendix A of~\cite{Fariborz:1999gr}, which assumes the SU(3) representation of the scalar mesons defined by Eqs.~(1.2), (1.3) from that work. Ref.~\cite{Aloni:2018vki}, using the same reference, provided the ALP decay matrix elements in terms of SU(3) representation of the scalar mesons that directly contradicts the definitions (1.2), (1.3).

As a result of these differences, our prediction of the decays of ALPs coupled solely to gluons differs from the one presented in~\cite{Aloni:2018vki}. In particular, we have found a somewhat larger value of the mass where the ChPT width matches with the perturbative QCD width, $m_{a} = 2.3\text{ GeV}$. 

\section{\texttt{Mathematica} notebook}
\label{app:mathematica-notebook}

To calculate and summarize the ALP production and decay rates, we implement the Lagrangian~\eqref{eq:lagr-chpt-1},~\eqref{eq:lagr-chpt-2}, as well as the RG flow for the couplings $\{c_{q}\} = c_{u,d,s,c,b,t}$, $c_{\ell} = c_{e,\mu,\tau}$ in a \texttt{Mathematica} notebook\footnote{Available on~\href{https://github.com/maksymovchynnikov/ALPs-phenomenology}{https://github.com/maksymovchynnikov/ALPs-phenomenology}}. The structure of the notebook is as follows. First, we define several ALP models at a scale $\Lambda$, such as the ALPs with universal fermion and gluon couplings. Then, we solve the RG equations for the fermion couplings (both the diagonal and FCNC couplings) at various scales $\Lambda$ following~\cite{Bauer:2020jbp,Bauer:2021mvw}, and interpolate the solutions.

Next, we implement the ChPT Lagrangian~\eqref{eq:lagr-chpt-1},~\eqref{eq:lagr-chpt-2} keeping the arbitrary values of the couplings $c_{u,d,s}$, and $c_{G}$. We diagonalize the quadratic ChPT Lagrangian to get the mixing angles and the ALP interactions. Then, we define the Feynman rules for the obtained Lagrangian and compute the matrix elements and decay widths of the ALP decay processes listed in Sec.~\ref{sec:decays}. In the last step, we specify the model and the scale $\Lambda$ and insert the resulting couplings into the decay widths. The resulting tabulated widths are then exported.

Finally, we compute the total ALP production rates described in Sec.~\ref{sec:production}. For this, we again use the RG flow of the couplings and the pre-computed ALP production cross-section in the gluon fusion for the unit value of $c_{G}$. 

For ALPs coupled solely to $W$ and $B$ bosons, only the production rates are currently evaluated. We will fully implement these models in future versions of the notebook.

\end{document}